\documentclass[a4paper, letter, longauth]{aa}

\usepackage{natbib}
\usepackage{graphicx}
\usepackage{txfonts}
\usepackage[bookmarks=false, colorlinks=true, citecolor=blue, linkcolor=blue]{hyperref}
\usepackage{upgreek}

\newcommand{\Mgiv}{[\ion{Mg}{iv}]}
\newcommand{\Oiv}{[\ion{O}{iv}]}
\newcommand{\Arvi}{[\ion{Ar}{vi}]}
\newcommand{\Feii}{[\ion{Fe}{ii}]}
\newcommand{\Arii}{[\ion{Ar}{ii}]}

\newcommand{\JWST}{\textit{JWST}}
\def\micron{\hbox{\,$\upmu$m}}
\newcommand{\Lsun}{\hbox{$L_{\rm \odot}$}}
\newcommand{\Msun}{\hbox{$M_{\rm \odot}$}}
\newcommand{\Zsun}{\hbox{$Z_{\rm \odot}$}}

\bibpunct{(}{)}{;}{a}{}{,}

\titlerunning{Extended \Mgiv\ emission tracing shocks in starbursts}
\authorrunning{Pereira-Santaella et al.}

\begin{document}

\title{Extended high-ionization [\ion{Mg}{iv}] emission tracing widespread shocks in starbursts seen by \JWST\slash NIRSpec}

\author{Miguel~Pereira-Santaella\inst{\ref{inst1}}
\and Ismael~Garc\'ia-Bernete\inst{\ref{inst2}}
\and Eduardo~Gonz\'alez-Alfonso\inst{\ref{inst3}}
\and Almudena~Alonso-Herrero\inst{\ref{inst4}}
\and Luis~Colina\inst{\ref{inst5}}
\and Santiago~Garc\'ia-Burillo\inst{\ref{inst6}}
\and Dimitra~Rigopoulou\inst{\ref{inst2},\ref{inst7}}
\and Santiago~Arribas\inst{\ref{inst5}}
\and Michele~Perna\inst{\ref{inst5}}
}

\institute{Instituto de F\'isica Fundamental, CSIC, Calle Serrano 123, 28006 Madrid, Spain \\
\email{miguel.pereira@iff.csic.es}\label{inst1}
\and
Department of Physics, University of Oxford, Keble Road, Oxford OX1 3RH, UK\label{inst2}
\and
Universidad de Alcal\'a, Departamento de F\'isica y Matem\'aticas, Campus Universitario, 28871 Alcal\'a de Henares, Madrid, Spain\label{inst3}
\and
Centro de Astrobiolog\'ia (CAB), CSIC-INTA, Camino Bajo del Castillo s/n, E-28692 Villanueva de la Ca\~nada, Madrid, Spain\label{inst4}
\and
Centro de Astrobiolog\'ia (CAB), CSIC-INTA, Ctra de Torrej\'on a Ajalvir, km 4, 28850, Torrej\'on de Ardoz, Madrid, Spain\label{inst5}
\and
Observatorio Astron\'omico Nacional (OAN-IGN)-Observatorio de Madrid, Alfonso XII, 3, 28014, Madrid, Spain\label{inst6}
\and
School of Sciences, European University Cyprus, Diogenes street, Engomi, 1516 Nicosia, Cyprus\label{inst7}
}

\abstract{
We report the detection of extended ($>$0.5--1\,kpc) high-ionization \Mgiv\,4.487\micron\ (80\,eV) emission in four local luminous infrared galaxies observed with \JWST\slash NIRSpec. Excluding the nucleus and outflow of the Type 1 active galactic nucleus (AGN) in the sample, we find that the \Mgiv\ luminosity is well correlated with that of H recombination lines, which mainly trace star forming clumps in these objects, and that the \Arvi\,4.530\micron\ (75\,eV), usually seen in AGN, is undetected. On 100--400\,pc scales, the \Mgiv\ line profiles are broader ($\sigma(\Mgiv)=90\pm 25$\,km\,s$^{-1}$) and shifted ($\Delta v$ up to $\pm$50\,km\,s$^{-1}$) compared to those of the H recombination lines and lower ionization transitions (e.g., $\sigma($Hu-12$)=57\pm 15$\,km\,s$^{-1}$).
The \Mgiv\ {kinematics follow the large scale rotating velocity field of these galaxies and the broad \Mgiv\ profiles are compatible with the broad wings detected in the H recombination lines.}
Based on these observational results, extended highly ionized gas more turbulent than the ambient interstellar medium, possibly as a result of ionizing shocks associated with star-formation, is the most likely origin of the \Mgiv\ emission.
We also computed new grids of photoionization and shock models to investigate where the \Mgiv\ line originates. Shocks with velocities of 100--130\,km\,s$^{-1}$ reproduce the observed line ratios and the \Mgiv\ luminosity agrees with that expected from the mechanical energy released by supernove (SNe) in these regions. Therefore, these models support shocks induced by SNe as the origin of the \Mgiv\ line.
Future studies on the stellar feedback from SNe will benefit from the \Mgiv\ line that {is little affected by obscuration} and, in absence of an AGN, can only be produced by shocks due to its high ionization potential.
}

\keywords{Galaxies: active -- Galaxies: evolution -- {Galaxies: starburst} -- Infrared: ISM}

\maketitle

\section{Introduction}\label{sec:intro}

Feedback from active galactic nuclei (AGN) is an important element for the evolution of galaxies (e.g., \citealt{Weinberger2018}).
However, detecting deeply buried AGN is still difficult due to dust obscuration. 
This is particularly challenging in local luminous and ultra-luminous infrared (IR) galaxies (U\slash LIRGs; $L_{\rm IR}>10^{11}$\,$L_\odot$). Local U\slash LIRGs are mostly interacting\slash merging systems, so they represent a key phase in the evolution of galaxies (e.g., \citealt{Patton2020}), and their activity (AGN and star-formation) occurs in extremely dust-embedded environments.

{Activity diagnostics in the optical} fail in many U\slash LIRGs since they host the most obscured nuclei in the local Universe ($N_{\rm H}>10^{25}$\,cm$^{-2}$; e.g., \citealt{Falstad2021, GarciaBernete2022_CON}).
Indirect methods using mid-IR spectroscopy pointed out the presence of AGN in a large fraction, $\sim$70\%, of ULIRGs (e.g., \citealt{Nardini2009, Veilleux2009}). These mid-IR methods suggested a modest AGN contribution to the total luminosity. However, radio and sub-mm observations, which are insensitive to dust extinction, revealed compact nuclei exceeding the surface brightness density of a maximal starburst, hence likely AGN, dominating the {bolometric} luminosity of the majority of ULIRGs \citep{BarcosMunoz2017, Pereira2021, Hayashi2021}.
Therefore, the direct detection {of} these AGN remains elusive and its exact contribution to the total luminosity is still {debated}.

The reduced dust extinction in the near- and mid-IR spectral ranges, along with the superb sensitivity and angular resolution of the James Webb Space Telescope (\JWST), for the first time allows the search for high-ionization emission lines, which may eventually provide direct evidence of these elusive AGN in U\slash LIRGs.
Specifically, the \Mgiv\,4.487\micron\ line, with a high ionization potential (IP; 80\,eV), is not expected in gas photoionized by young stars since they do not emit significant quantities of photons with the energies beyond 54\,eV (He$^+$ ionization energy). Therefore, the detection of the \Mgiv\ line, which lies close to the minimum of the dust opacity (e.g., \citealt{Corrales2016}), has been proposed as an excellent tracer of deeply embedded AGN (e.g., \citealt{Satyapal2021}). However, fast ionizing shocks ($>$80\,km\,s$^{-1}$; \citealt{Sutherland2017}) {can also} produce highly ionized gas {that can emit high-ionization lines. For instance, the \Oiv\,25.9\micron\ line (IP 55\,eV), which is well correlated with the AGN power (e.g., \citealt{Diamond2009, Pereira2010c, GarciaBernete2016}), is also associated with shocks in starbursts (e.g., \citealt{Lutz1998, Petric2011, AAH2012a}). }

In this Letter, we investigate the origin (AGN, star-formation, or shocks) of the extended \Mgiv\ emission detected in a sample of four local LIRGs observed with the \JWST\slash Near Infrared Spectrograph (NIRSpec; \citealt{Jakobsen2022}).
We first analyze the spatial distribution of the \Mgiv\ emission and compare the \Mgiv\ line profiles with those of other low-ionization and H recombination lines. Then, we use new grids of photoionization and shock models to determine the physical and dynamical conditions of the gas producing this line.

\section{Sample and data reduction}\label{s:data}

Four local (39 < $D_{\rm L}$\slash Mpc < 161) LIRGs, with $L_{\rm IR}$=10$^{11.6-11.9}$\Lsun, were observed with \JWST\slash NIRSpec using the high spectral resolution grating G395H ($R$$\sim$1900--3600; 2.87--5.27\micron) and the integral field unit (IFU) mode \citep{Boker2022}, and with \JWST\slash Mid-Infrared Instrument (MIRI) using all the bands of the Medium Resolution Spectrograph (MRS; \citealt{Wright2023, Argyriou2023}).
These observations were part of the Director's Discretionary Early Release Science (DD-ERS) Program \#1328 (PI: L.~Armus and A.~Evans).
These four objects cover a wide range of properties: from pre-mergers to late merger stages and from heavily obscured AGN to Type 1 AGN (see Table~\ref{tbl_sample}).

\begin{table}
\caption{Sample of local LIRGs}
\label{tbl_sample}
\centering
\begin{small}
\begin{tabular}{lcccccc}
\hline \hline
\\
Object & $D_{\rm L}$\,$^a$ & log\,$L_{\rm IR}\slash L_\odot$\,$^a$ & Interaction & AGN\,$^c$ \\
& (Mpc) & & stage\,$^b$ &  \\
\hline
VV~114~E & 85.5 & 11.71 & c & Obs. ? (1,2)\\ %
NGC~3256~N\,$^*$ & 38.9 & 11.64 & d & No (3) \\ %
NGC~3256~S\,$^*$ & & &  & Obs. (4) \\
II~Zw~096 & 161 & 11.94 & c & Obs. ? (5) \\  %
NGC~7469 & 70.8 & 11.65 & a & Type 1 (6) \\ %
\hline
\end{tabular}
\end{small}
\tablefoot{
$^{(a)}$ Luminosity distance and 8-1000\micron\ IR luminosity from \citet{Armus09}. 
$^{(b)}$ Interaction stage (a=pre-merger, c=mid-stage merger, and d=late stage merger) from \citet{Stierwalt2013}.
$^{(c)}$ AGN classification (Obs. = dust obscured AGN based on indirect evidences). ``?'' indicates inconclusive AGN detections.
$^{(*)}$ {The separation between the two nuclei of NGC~3256 is $\sim$5\farcs2, thus their NIRSpec observations ($\sim$0\farcs16 angular resolution at 4.5\micron) are well spatially resolved.}
}
\tablebib{(1) \citealt{Rich2023}; (2) \citealt{Donnan2023}; (3) \citealt{Pereira2010}; (4) \citealt{Ohyama2015}; (5) \citealt{Inami2022}; (6) \citealt{Osterbrock1993}}
\end{table}

For the data reduction, we used the \JWST\ calibration pipeline (version 1.12.4; \citealt{Bushouse_1_12_4}) and the context 1197. We followed the standard reduction recipe complemented by a number of custom steps to reduce the effect of bad pixels and cosmic rays. Further details on the data reduction of the NIRSpec and MIRI\slash MRS data observations be found in \citet{Pereira2022, Pereira2024} and \citet{GarciaBernete2024_ice}.

\section{Analysis and results}\label{s:results}

\subsection{Spatially extended \Mgiv\ emission}\label{s:maps}

\begin{figure}
\centering
\includegraphics[width=0.4\textwidth]{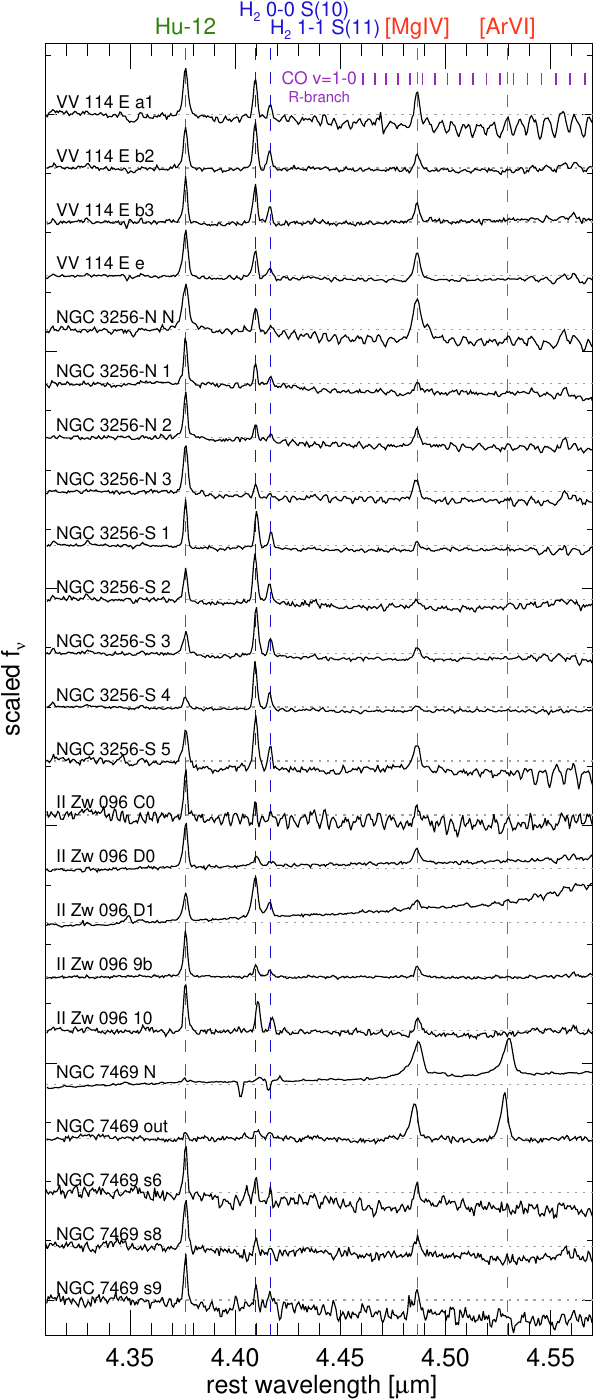}
\caption{Scaled and shifted \JWST\slash NIRSpec 4.31--4.57\micron\ spectra of the selected regions (see Sect.~\ref{s:maps} and Figs.~\ref{fig_map1} and \ref{fig_map2}-\ref{fig_map5}). The wavelength of the transitions tracing ionized gas (Hu-12), warm molecular gas (H$_2$ 0--0 S(10)\,4.410\micron\ and \hbox{1--1} S(11)\,4.417\micron), highly ionized gas (\Mgiv\ and \Arvi), and the CO $v$=1--0 band are indicated by the green, blue, red, and purple vertical lines, respectively.
\label{fig_spectra}}
\end{figure}

\begin{figure*}
\centering
\includegraphics[width=0.92\textwidth]{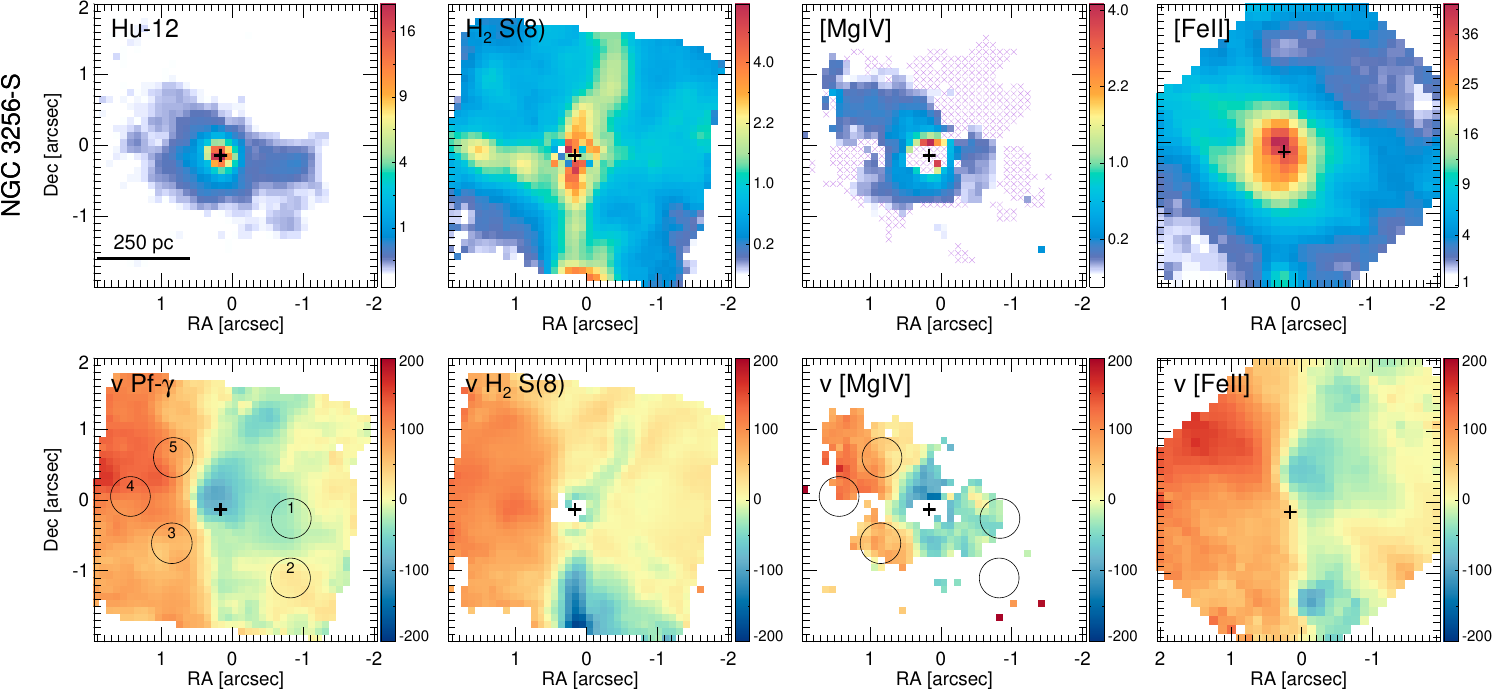}
\caption{Line maps and velocity fields for NGC~3256~S. The remaining objects are in Figs.~\ref{fig_map2}-\ref{fig_map5}. {\it Top row from left to right}: Line maps of Hu-12, H$_2$ S(8), \Mgiv, and \Feii. The areas filled with purple crosses in the \Mgiv\ panel correspond to regions where the CO \hbox{$v$=1--0} band is {strong and complicates the modeling of the} \Mgiv\ line. The black cross marks the position of the nucleus.
{\it Bottom row from left to right}: Velocity fields from single Gaussian fits of Pf-$\gamma$, H$_2$ S(8), and \Mgiv. The circles in the first and third panel mark the location and size of the selected regions.
The units of the color scale are 10$^{-15}$\,erg\,cm$^{-2}$\,s$^{-1}$\,arcsec$^{-1}$ for the line maps and km\,s$^{-1}$ for the velocity fields.
\label{fig_map1}}
\end{figure*}

Fig.~\ref{fig_spectra} shows the rest-frame 4.31--4.57\micron\ spectra from 23 regions selected in the four systems. To extract these spectra we used apertures with a diameter of 0\farcs54 (100--400\,pc depending on the target). We did not apply any aperture correction as the emission appears spatially resolved.
Nevertheless, in a worst-case scenario of a point source, the error in the line ratios discussed below (Sect.~\ref{s:models}) would be small, $<$5\%.
The regions were selected to cover the extended \Mgiv\ emission detected in the line maps (see Figs.~\ref{fig_map1} and \ref{fig_map2}-\ref{fig_map5}).
For simplicity, in this Letter we excluded regions with strong CO \hbox{$v$=1--0} band, since the \Mgiv\ transition is almost coincident with the {CO \hbox{$v$=1--0}} R(24) 4.489\micron\ line (Fig.~\ref{fig_spectra}). We note, however, that after modeling the band, it is possible to detect \Mgiv\ emission also in these regions (see \citealt{GonzalezAlfonso2024, GarciaBernete2024_SODA}).

The spectra show clear \Mgiv\ detections, two H$_2$ lines, the Humphreys-12~4.376\micron\ (Hu-12) recombination line, and, in some cases, weak absorptions from the R-branch of the CO \hbox{$v$=1--0} band. The \Arvi\,4.530\micron\ transition, with an IP of 75\,eV slightly lower than \Mgiv, is only detected in the nucleus and outflow of the Type 1 AGN NGC~7469 with both lines having comparable intensities (see also \citealt{Bianchin2023}).
The latter is consistent with the {\it Infrared Space Observatory} ({\it ISO}) detections of the \Mgiv\ and \Arvi\ line pair in two AGN (Circinus and NGC~1068), where the \Arvi\ line is $\sim$2 times brighter \citep{Sturm2002}. Based on these previous results for AGN, the absence of the \Arvi\ transition in the remaining regions is surprising and suggests that distinct physical conditions, maybe not related to an AGN, can lead to this \Mgiv\ emission.

\begin{figure}
\centering
\includegraphics[width=0.33\textwidth]{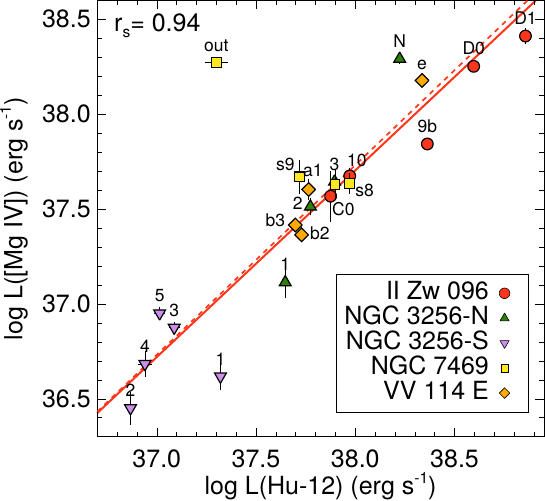}
\includegraphics[width=0.33\textwidth]{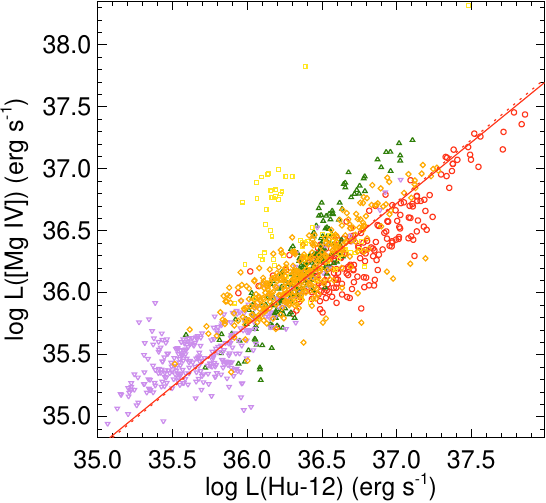}
\caption{\Mgiv\ vs. Hu-12 luminosities {of the selected regions (top) and for every spaxel (bottom).}
Different colors and symbols are used to identify the regions of each object. {In the top panel,} the labels of the points are the names of the regions. The Spearman correlation coefficient, $r_s$, is indicated. The solid {and dashed} red lines are the best fit ($\log y = 0.30 + 0.98 \log x$) and best linear fit ($\log y = (-0.29 \pm 0.18) + \log x$), {respectively, to the luminosities of the regions}.
\label{fig_flux_relation_hu12}}
\end{figure}

{To determine the origin of the \Mgiv\ line, we created 2D emission maps of tracers of the ionized gas phase (Hu-12 and Pf-$\gamma$\,3.741\micron), the warm molecular gas phase (H$_2$ \hbox{0--0} S(8)~5.053\micron), and fast shocks (\Feii\,5.340\micron).}
{Warm molecular gas excited by shocks has been identified in interacting systems and U\slash LIRGs (e.g., \citealt{Guillard2009, Pereira2014}) but the shock velocities are typically lower than that of the ionizing shocks traced by \Feii.}
We extracted the spectra using a running 2x2 spaxels box and fitted a local linear continuum level and a Gaussian profile to determine the flux and velocity at each position (Figs.~\ref{fig_map1} and \ref{fig_map2}-\ref{fig_map5}). The \Mgiv\ morphology is spatially extended, from few hundreds pc to $\sim$1\,kpc, and its spatial distribution is more similar to that of the ionized and shocked gas (Hu-12 and \Feii) than to the warm molecular H$_2$ emission. Actually, for the regions selected before, there is a very good linear correlation between the Hu-12 and the \Mgiv\ luminosities (excluding the nucleus and outflow of NGC~7469) which spans almost 2\,dex and has a Spearman correlation coefficient of 0.94 (Fig.~\ref{fig_flux_relation_hu12} {top}). {This correlation also holds when we consider the individual spaxels (Fig.~\ref{fig_flux_relation_hu12} bottom). The correlation between the shock tracer \Feii\ and \Mgiv\ is also good (Fig.~\ref{fig_flux_relation_h2} {bottom panel}).}
For the warm H$_2$, the correlation coefficient is lower, 0.72, and the correlation, if real, would not be linear (Fig.~\ref{fig_flux_relation_h2} {top panel}).

\subsection{Broad \Mgiv\ line profiles}\label{s:profiles}

\begin{figure*}
\centering
\includegraphics[width=\textwidth]{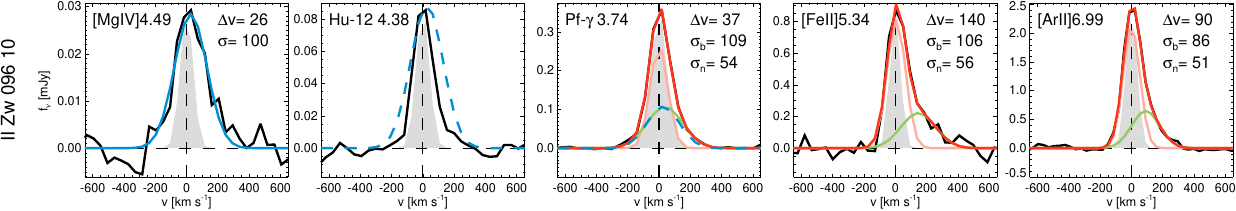}
\includegraphics[width=\textwidth]{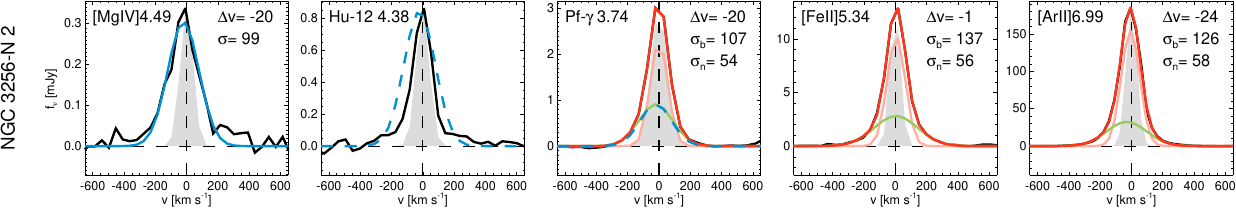}
\caption{Observed profiles of the \Mgiv, Hu-12, Pf-$\gamma$, \Feii, and \Arii\ emission lines (solid black line) for two selected regions. The remaining regions are shown in Fig.~\ref{fig_prof_apx}.
The filled gray Gaussian represents the instrumental spectral resolution.
The blue line in the first panel is the best single Gaussian fit to the \Mgiv\ profile. The \Mgiv\ model is also plotted {as a blue dashed line} in the second and third panels, normalized to the peak of Hu-12 and the broad Pf-$\gamma$ component, respectively, for reference.
The two component fits for Pf-$\gamma$, \Feii, and \Arii, are represented by the pink (narrow component) and green (broad component) lines and the total model in red. For reference, the dashed lines mark the rest velocity for each region derived from the narrow component of Pf-$\gamma$ and the zero flux level.
The observed (not corrected for instrumental resolution) $\sigma$ and velocity shift derived from the Gaussian fits are indicated in each panel in km\,s$^{-1}$.
\label{fig_profiles}}
\end{figure*}

\begin{figure}[h!]
\centering
\includegraphics[width=0.3\textwidth]{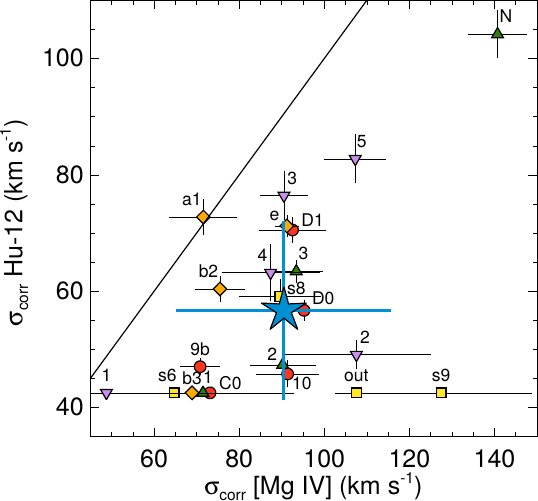}\\[0.3cm]
\includegraphics[width=0.3\textwidth]{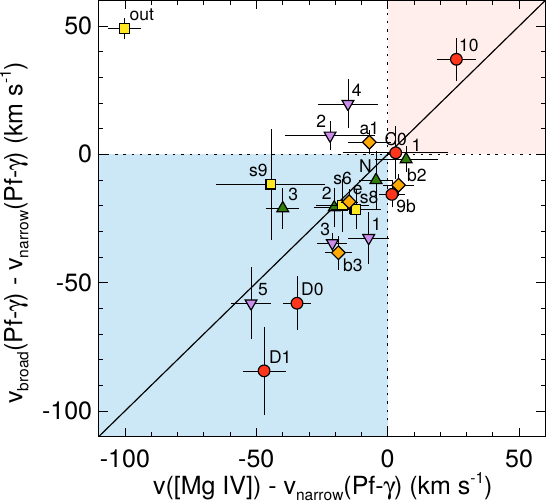}
\caption{Comparison between the width, corrected for the instrumental resolution, of the \Mgiv\ and Hu-12 lines (top) and between the velocity shifts of \Mgiv\ and the broad Pf-$\gamma$ component relative to the narrow \hbox{Pf-$\gamma$} component (bottom). The symbols are as in Fig.~\ref{fig_flux_relation_hu12}. The large blue star in the top panel represents the mean $\sigma_{\rm corr}$ in each axis.
The solid black line is the one-to-one relation.
The shaded blue and red areas in the bottom panel indicate blue and red velocity shifts, respectively.
\label{fig_kin_relations}}
\end{figure}

We compared the kinematics of the gas emitting the \Mgiv\ line with that of the ionized gas traced by other transitions by modeling their profiles.
In particular, we analyzed the following lines in the spectra of the selected regions: the Hu-12 line that is close in wavelength to \Mgiv\ and the effects of differential extinction are minimized;
Pf-$\gamma$, which can be more affected by extinction in these dusty galaxies ($A_{{\rm Pf}-\gamma}\slash A_{\rm \Mgiv}$= 1.14; \citealt{Chiar2006}), but is intrinsically brighter than Hu-12 (Pf-$\gamma$\slash Hu-12 = 5.53 at 10\,000\,K; \citealt{Storey1995});
\Feii\,5.340\micron, which is a good tracer of shocks \citep{Koo2016}; and \Arii\,6.985\micron, which also traces ionized gas, but has significantly higher signal to noise ratios (SNR).

The observed line profiles and best-fit models are presented in Figs.~\ref{fig_profiles} and \ref{fig_prof_apx}. A single Gaussian component reproduces well the \Mgiv\ and Hu-12 profiles. We find that the \Mgiv\ profile is systematically broader than the Hu-12 profile (top panel of Fig.~\ref{fig_kin_relations}). The average widths, corrected for the instrumental resolution, are: $\sigma_{\rm corr}(\Mgiv)=90\pm 25$\,km\,s$^{-1}$ and $\sigma_{\rm corr}({\rm Hu-12})=57\pm 15$\,km\,s$^{-1}$. In addition, the \Mgiv\ emission presents velocity shifts, up to $\pm$50\,km\,s$^{-1}$, with respect to Pf-$\gamma$ {(the NIRSpec absolute wavelength accuracy for the G395H grating is $\sim$12\,km\,s$^{-1}$; priv. comm.).}
In the higher SNR profiles of Pf-$\gamma$, a second broader component is clearly detected through higher velocity wings in all the profiles. Wings are also detected in the \Feii\ and \Arii\ lines (Fig.~\ref{fig_profiles}, two right most panels). Broad profiles ($>$200\,km\,s$^{-1}$) have been detected in the mid-IR lines of local U\slash LIRGs typically associated with the narrow line region (NLR) of AGN and ionized outflows \citep{Spoon2009, Dasyra2011, AAH2013, U2022, GarciaBernete2022, Armus2023}. However, we note that the regions selected here are widespread over the entire extent of these objects and not specifically selected to encompass outflows or the NLR of AGN.

We show the best-fit two Gaussian model for these lines in the third, fourth, and fifth columns of Figs.~\ref{fig_profiles} and \ref{fig_prof_apx}. We find a remarkable similarity between the widths and velocity shifts of the \Mgiv\ and the broad Pf-$\gamma$ profiles. The
$\sigma^{\rm broad}_{\rm corr}$(\hbox{Pf-$\gamma)$}\,$=105\pm 15$\,km\,s$^{-1}$ is comparable to $\sigma_{\rm corr}(\Mgiv)$ and the great majority (85\%) of the velocity shifts have the same sign (i.e., lie in the blue- or red- shaded areas in the bottom panel of Fig.~\ref{fig_kin_relations}).

In most of the cases ($\sim$75\%) the \Mgiv\ line is blue-shifted relative to the narrow component in Pf-$\gamma$.
Although extinction is greatly reduced at $\sim$4.5\micron, it might affect the receding side of the \Mgiv\ emission and explain this uneven distribution of the sign of the shifts. A detailed study of the extinction is beyond the scope of this Letter. We also note that the sign of the shifts of the broad component of \Feii\ and \Arii\ follow the same trend (Fig.~\ref{fig_kin_relations_fear}) and the widths are also comparable to that of the \Mgiv\ line, $\sigma^{\rm broad}_{\rm corr}$(\Feii)=$112\pm 27$\,km\,s$^{-1}$ and $\sigma^{\rm broad}_{\rm corr}$(\Arii)=$102\pm 20$\,km\,s$^{-1}$.

\begin{figure*}
\centering
\includegraphics[width=0.82\textwidth]{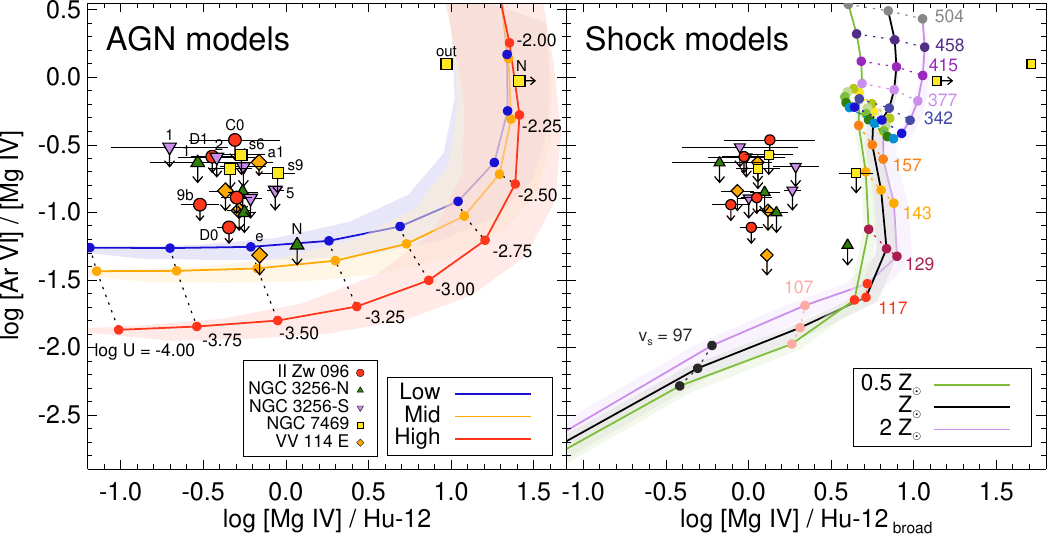}
\caption{\Arvi\slash\Mgiv\ vs. \Mgiv\slash Hu-12 ratio predicted by AGN (left) and shock (right) models. {For the shock models we only consider the flux of the broad component of Hu-12 estimated from the broad to narrow flux ratio of Pf-$\gamma$.} {\it Left:} The blue, orange, and red solid lines mark the reference AGN models ($Z_\odot$ and $n_{\rm H}=10^3$\,cm$^{-3}$) for the low, mid, and high Eddington ratio SEDs, respectively.
The shaded areas highlight the range of ratios predicted varying the metallicities (0.4--2\,$Z_\odot$) and $n_{\rm H}$ (10$^2$--10$^5$\,cm$^{-3}$). Models with the same ionization parameter, $U$, are connected by dashed lines.
{\it Right:} The green, black, and purple solid lines mark the reference shock models (ram pressure $R=10^6$) for 0.5$Z_\odot$, $Z_\odot$, and 2$Z_\odot$ metallicities, respectively.
The shaded areas highlight the range of ratios predicted varying $R$ (10$^4$--10$^8$).
Models with the same shock velocity, $v_{\rm s}$, are connected by dashed lines.
The observed ratios use the same symbols as in Fig.~\ref{fig_flux_relation_hu12}.
The models are described in Sect.~\ref{s:models} and Appendix~\ref{apx:models}.
\label{fig_model_agn_shock}}
\end{figure*}

\section{Origin of the \Mgiv\ emission: star-formation, AGN, and shock models}\label{s:models}

Despite its high IP, the results described in Sect.~\ref{s:results} are not fully consistent with an AGN origin of the \Mgiv\ emission because: (1) it is extended and correlated with the Hu-12 emission, a H recombination line tracing star forming regions based on its morphology; (2) the \Arvi\ line, with a slightly lower IP and typically found in other AGN, is undetected, except in the outflow and nucleus of the Type 1 AGN; and (3) the \Mgiv\ profile is broader and shifted compared to the other transitions.
Therefore, we created three grids of models (star-formation, AGN, and shocks) to investigate the origin of the \Mgiv\ emission.
We briefly describe the models in this section and provide further details in Appendix~\ref{apx:models}.

We modeled the emission from gas photoionized by AGN or star-formation using the spectral synthesis code \textsc{Cloudy} \citep{Chatzikos2023}. We mostly followed \citet{Pereira2017FIR} to create these grids but updated the spectral energy distributions (SEDs) of the incident radiation field. For the star-formation grids, we used the Binary Population and Spectral Synthesis (BPASS) library \citep{Eldridge2017}, which includes stripped-envelope stars that have harder ionizing spectra (e.g., \citealt{Gotberg2019}) and could contribute to the \Mgiv\ emission.
For the AGN models, we used the three SEDs for low, medium, and high Eddington ratios derived by \citet{Jin2012}, which provide a more realistic approximation to the intrinsic SED of AGN than a power-law.

Based on these grids, we can reject that the \Mgiv\ line originates in gas photoionized by stars since the observed \Mgiv\slash Hu-12 ratio ($\sim$$0.5$; Fig.~\ref{fig_flux_relation_hu12}) is $>$2.3\,dex larger than the predicted ratio (Fig.~\ref{fig_sf_model}).

The AGN photoionization predictions for the \Mgiv\slash Hu-12 and \Arvi\slash\Mgiv\ ratios are shown in the left panel of Fig.~\ref{fig_model_agn_shock}. The outflow and nucleus of the Sy1 NGC~7469 lie close to the grid with ionization parameters, $\log U$, between $-2.00$ and $-2.25$ in the range of Seyfert AGN (e.g., \citealt{PerezDiaz2022}). For the remaining regions, the lower \Mgiv\slash Hu-12 ratio and the non-detection of \Arvi\ would imply $\log U<-3.25$, which is low but is also found in some AGN and low-ionization nuclear emission regions (LINERs; e.g., \citealt{Thomas2018}).

Finally, we created a grid of shock models using \textsc{MAPPINGS V} following \citet{Sutherland2017}, which includes several improvements compared to the models in \citet{Allen2008} (see Appendix~\ref{apx:mod_shock}). The {\Mgiv\slash Hu-12$_{\rm broad}$} ratio and {the} \Arvi\ {3$\sigma$} upper limit are compatible with the emission produced by shocks with shock velocities $v_{\rm s}=100-130$\,km\,s$^{-1}$ (Fig.~\ref{fig_model_agn_shock} right). The $v_{\rm s}$ is not directly equivalent to $\sigma_{\rm corr}$ \hbox{(e.g., \citealt{Ho2014, Perna2020})}, however, the fact that they are similar, $\sigma_{\rm corr}(\Mgiv)=90\pm 25$\,km\,s$^{-1}$ (Sect.~\ref{s:profiles}), supports that shocks are a possible origin of the \Mgiv\ emission.

In summary, while we cannot completely reject the AGN origin {in a very low $\log U$ environment} based on these line ratios, the {morphology} and kinematic properties of the \Mgiv\ emission (broader profile and velocity shifts) favor the shock origin.
Under this assumption, the good correlation between the extended \Mgiv\ emission and the recombination line Hu-12, which traces star forming clumps, is naturally explained if the shocks producing the \Mgiv\ emissions are powered by supernove (SNe).
To estimate if the energy released by SNe can produce the observed \Mgiv\ luminosity relative to the star-formation rate (i.e., \Mgiv\slash Hu-12 ratio), we combined stellar population evolution models and the shock models. 
Shock models predict efficient \Mgiv\ emission at $v_{\rm s}$$\sim$110--160\,km\,s$^{-1}$, so observing \Mgiv\ emission at these velocities is also favored (Fig.~\ref{fig_shock_energy}). Using these relations (see Appendix~\ref{apx:mod_shocks_energy}), the predicted log\,\Mgiv\slash Hu-12 ratio from a star forming region with shocks produced by SNe is between $-1.4$ and $-0.5$.
The observed log\,\Mgiv\slash  \hbox{Hu-12} ratio, $-0.29\pm 0.18$, lies at the upper end of the range but is consistent with it considering the relatively large uncertainties of both the shock models (e.g., $v_{\rm s}$, Mg abundance) and the star-formation models (e.g., initial mass function, energy per SN, electron temperature). Therefore, it is plausible that SNe can provide enough mechanical energy to generate the observed \Mgiv\ emission.

\section{Summary and conclusions}\label{s:summary}

Using \JWST\slash NIRSpec integral field spectroscopy, we investigated the extended high-ionization \Mgiv\,4.487\micron\ (IP 80\,eV) emission in a sample of four local LIRGs, only one of them hosting a Seyfert-like AGN.
Excluding the nucleus and outflow of this AGN, we find that shocks related to star formation are the most likely origin of the extended \Mgiv\ emission since: (1) the \Mgiv\ luminosity is well correlated with the recombination line Hu-12~4.376\micron, which traces star forming clumps in these objects; {and} the \Arvi\,4.530\micron\ line (IP 75\,eV), which is common in the spectra of AGN, remains undetected; and (2) the \Mgiv\ profile is broader ($\sigma_{\rm corr}(\Mgiv)=90\pm 25$\,km\,s$^{-1}$) and shifted up to $\pm$50\,km\,s$^{-1}$ relative to the recombination lines ($\sigma_{\rm corr}($Hu-12$)=57\pm 15$\,km\,s$^{-1}$). The \Mgiv\ line kinematics actually track the faint wings of the lower ionization lines (Pf-$\gamma$\,3.741\micron, \Feii\,5.340\micron, and \Arii\,6.985\micron) and {resembles} the large scale rotating velocity field of the galaxies.

Supporting this interpretation, shock models with $v_{\rm s}\sim$100--130\,km\,s$^{-1}$ (i.e., similar to the $\sigma$ of \Mgiv\ and the broad component in other species) are consistent with the observed \Mgiv\slash Hu-12 ratio and the \Arvi\ upper limit. Also, the \Mgiv\ luminosity is comparable to that expected from shocks associated with the mechanical energy released by SNe in these regions.
{Based on these results, the detection of \Mgiv\ emission is expected in pure star-forming galaxies, at least in objects with high gas and star-formation surface densities similar to those of local LIRGs (e.g., \citealt{SanchezGarcia2022}).}

{Due to its high IP, \Mgiv\ offers a unique view of the shocked gas by tracing the highly ionized phase, contrasting with classic shock tracers such as the \Feii\ transitions, which have a lower IP of 7.9\,eV and are more easily ionized.}
Compared to the \Oiv\,25.9\micron\ line (IP 55\,eV), the higher sensitivity and angular and spectral resolutions of \JWST\ at the shorter wavelength of the \Mgiv\ line will be beneficial for future studies of stellar feedback caused by these widespread shocks in strong starbursts.

\begin{acknowledgements}
{We thank the referee for their useful comments and suggestions.}
The authors acknowledge the GOALS ERS team for developing their observing program.
MPS acknowledges support from grant RYC2021-033094-I funded by MICIU/AEI/10.13039/501100011033 and the European Union NextGenerationEU/PRTR.
{IGB and DR acknowledge} support from STFC through grants ST/S000488/1 and ST/W000903/1.
{AAH} acknowledges support from grant PID2021-124665NB-I00 funded by the Spanish
Ministry of Science and Innovation and the State Agency of Research
MCIN/AEI/10.13039/501100011033  and ERDF A way of making Europe.

This work is based on observations made with the NASA/ESA/CSA James Webb Space Telescope. The data were obtained from the Mikulski Archive for Space Telescopes at the Space Telescope Science Institute, which is operated by the Association of Universities for Research in Astronomy, Inc., under NASA contract NAS 5-03127 for JWST; and from the European JWST archive (eJWST) operated by the ESAC Science Data Centre (ESDC) of the European Space Agency. These observations are associated with program \#1328.

\end{acknowledgements}

\bibliographystyle{aa}

\appendix
\onecolumn

\section{Emission line maps}

\begin{figure}[h]
\centering
\includegraphics[width=0.92\textwidth]{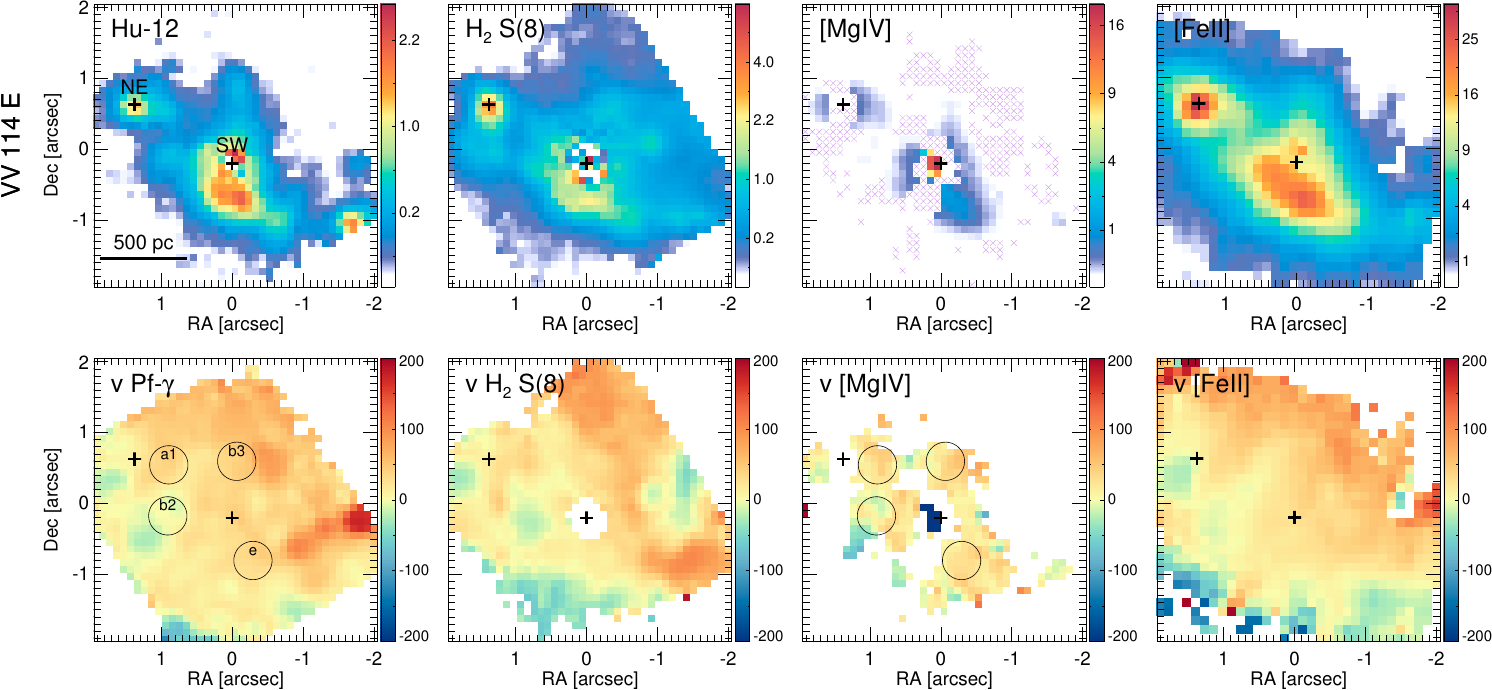}
\caption{Same as Fig.~\ref{fig_map1} but for VV~114~E.
\label{fig_map2}}
\end{figure}

\begin{figure}[h]
\centering
\includegraphics[width=0.92\textwidth]{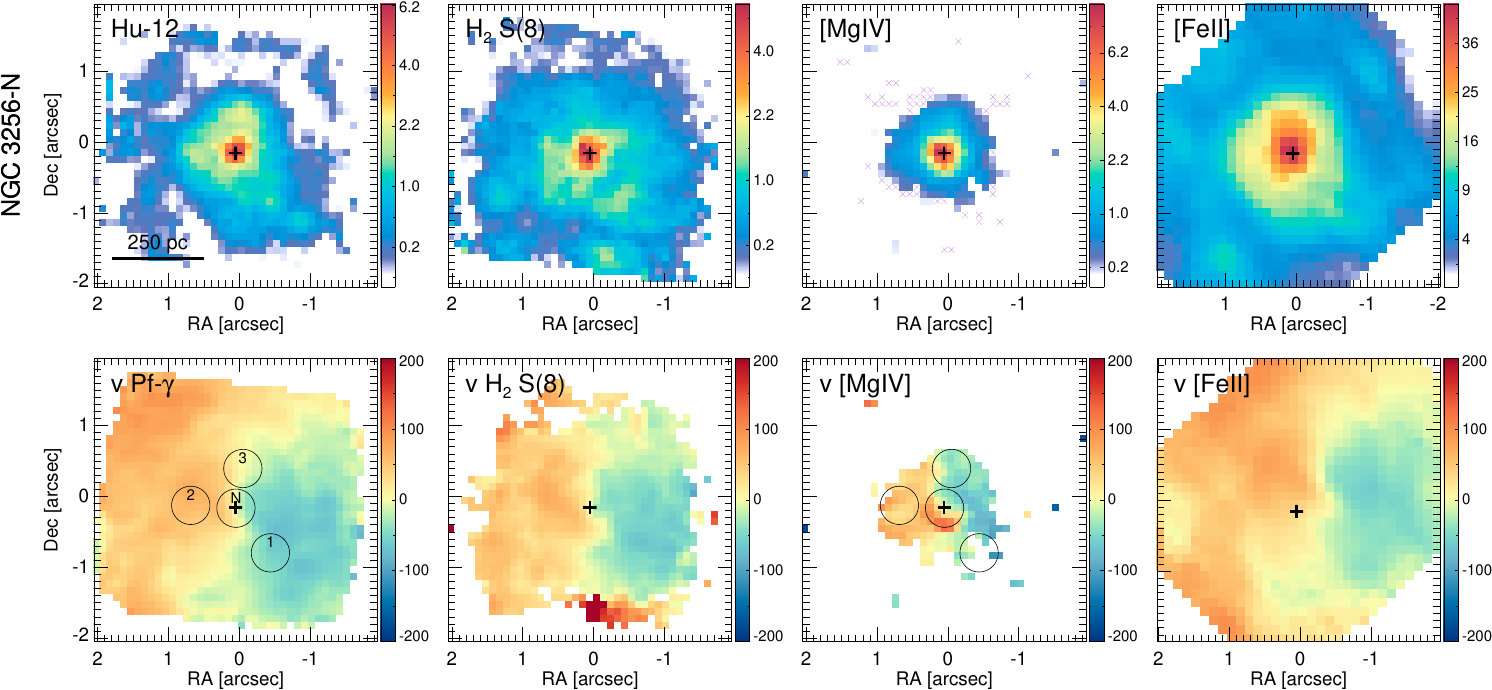}
\caption{Same as Fig.~\ref{fig_map1} but for NGC~3256~N.
\label{fig_map3}}
\end{figure}

\begin{figure}[h]
\centering
\includegraphics[width=0.92\textwidth]{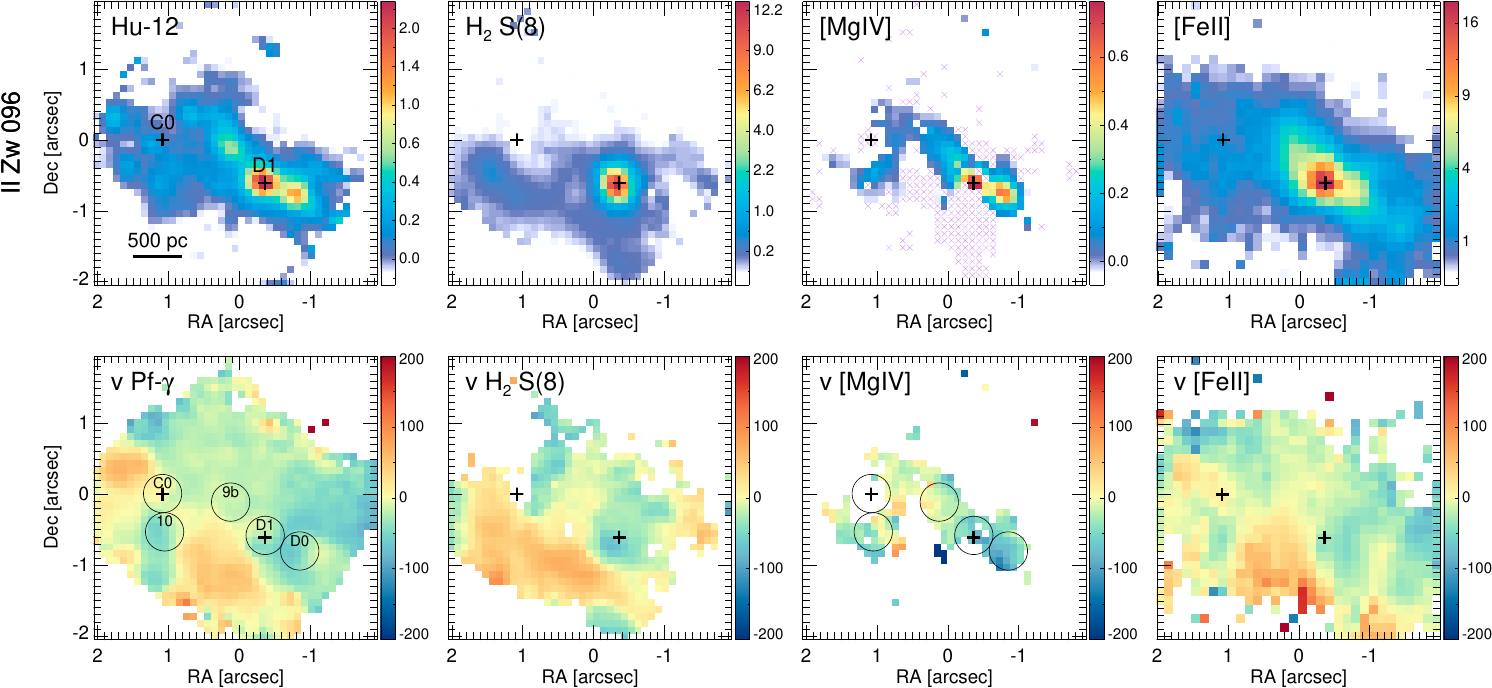}
\caption{Same as Fig.~\ref{fig_map1} but for II~Zw~096.
\label{fig_map4}}
\end{figure}

Figs.~\ref{fig_map2}-\ref{fig_map5} show the emission maps for the complete sample.

\clearpage

\begin{figure}[h]
\centering
\includegraphics[width=0.92\textwidth]{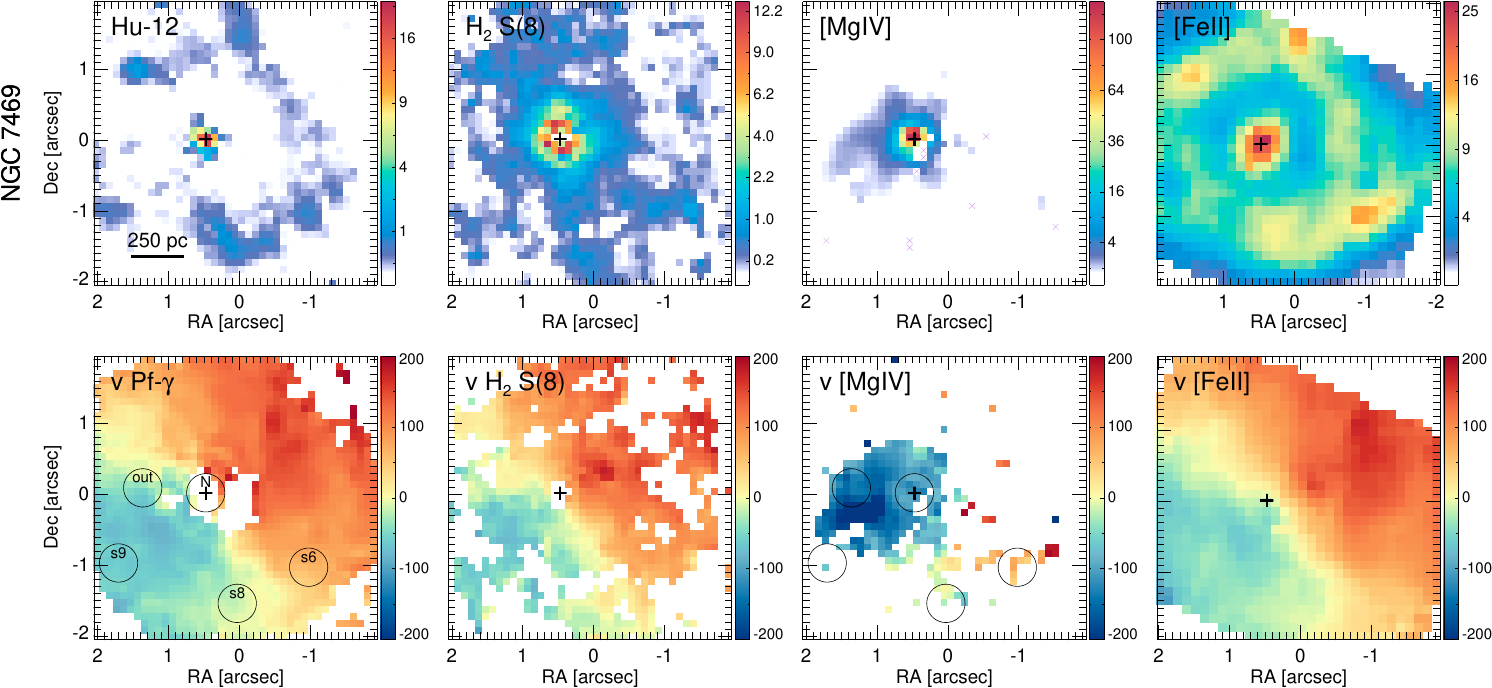}
\caption{Same as Fig.~\ref{fig_map1} but for NGC~7469.
\label{fig_map5}}
\end{figure}

\clearpage

\section{Emission line profiles}

Fig.~\ref{fig_prof_apx} shows the emission line profiles for all the selected regions.

\begin{figure}[h]
\centering
\includegraphics[width=0.97\textwidth]{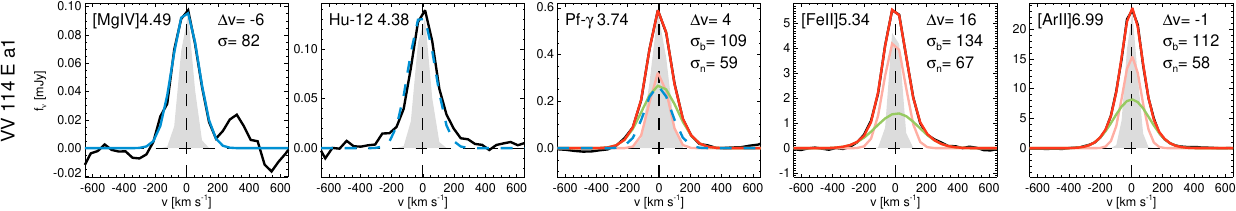}
\includegraphics[width=0.97\textwidth]{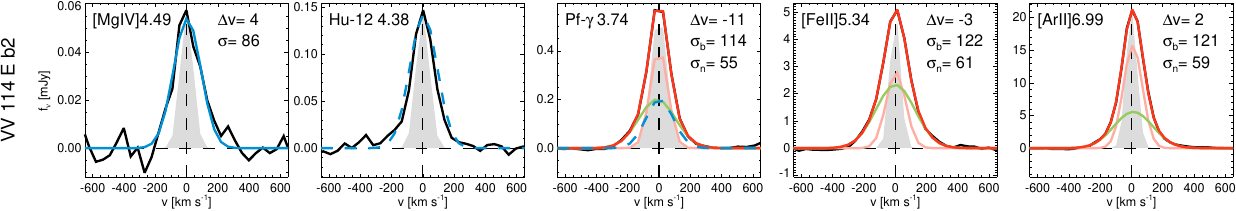}
\includegraphics[width=0.97\textwidth]{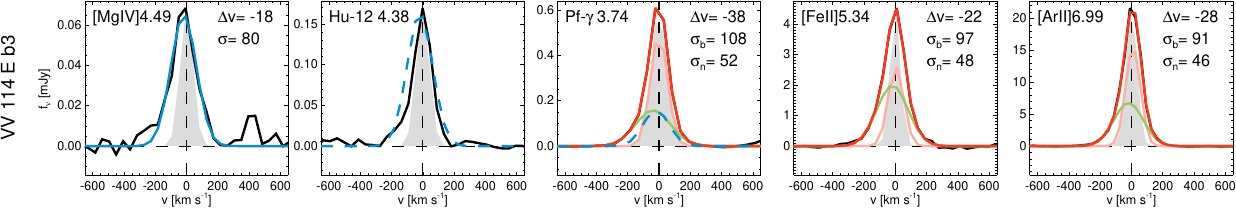}
\includegraphics[width=0.97\textwidth]{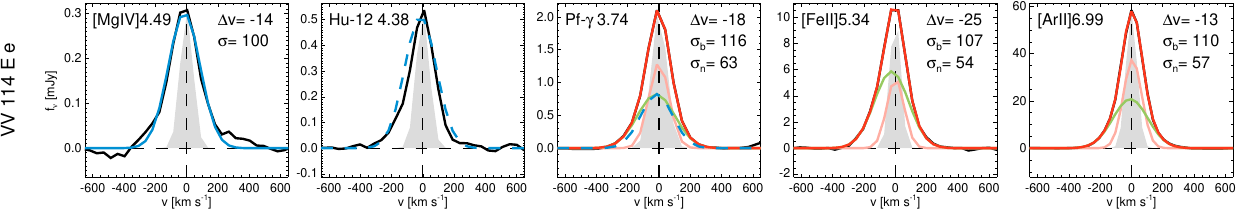}
\includegraphics[width=0.97\textwidth]{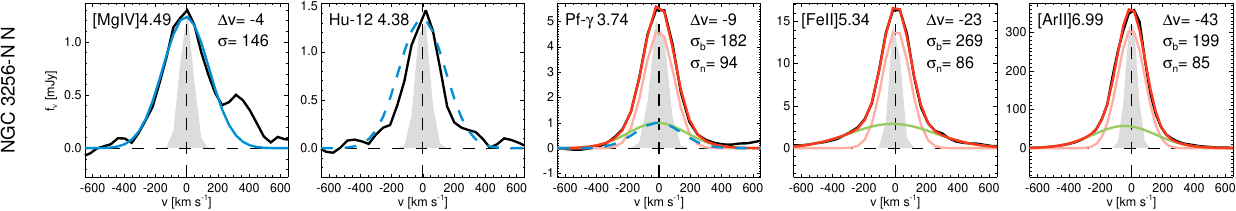}
\includegraphics[width=0.97\textwidth]{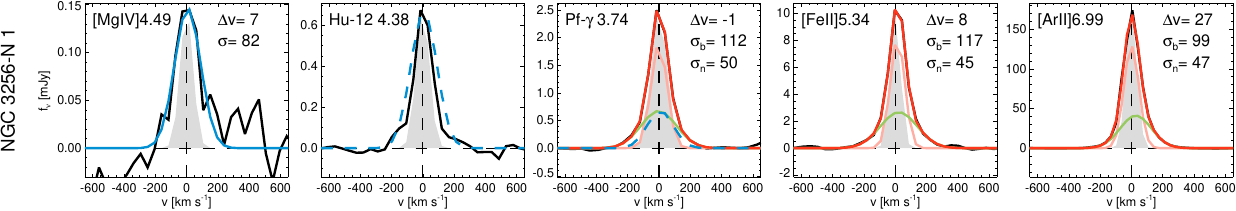}
\includegraphics[width=0.97\textwidth]{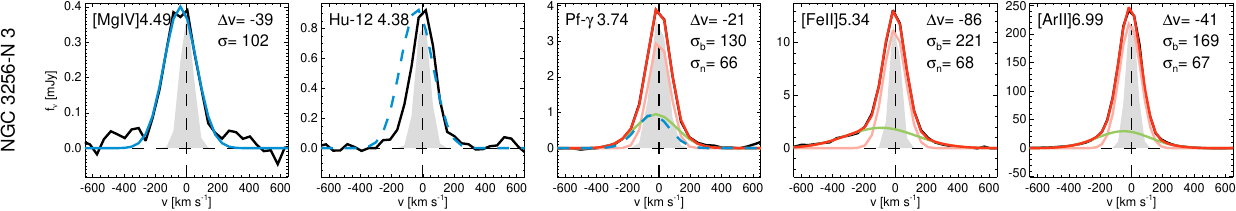}
\caption{Continued.\label{fig_prof_apx}}
\end{figure}

\begin{figure}[h]
\addtocounter{figure}{-1}
\centering
\includegraphics[width=0.97\textwidth]{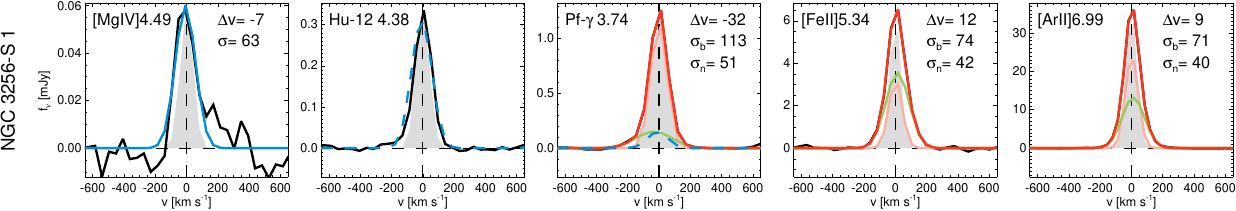}
\includegraphics[width=0.97\textwidth]{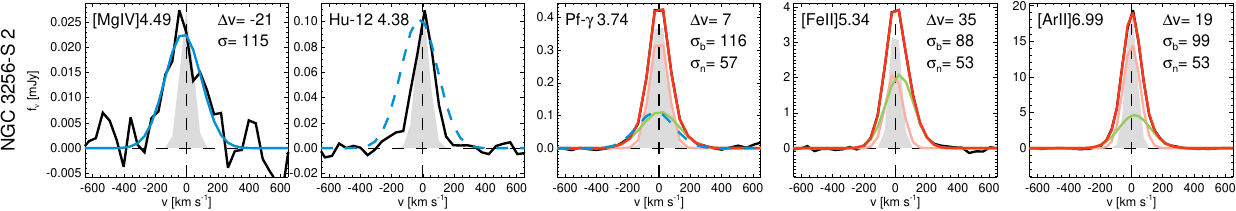}
\includegraphics[width=0.97\textwidth]{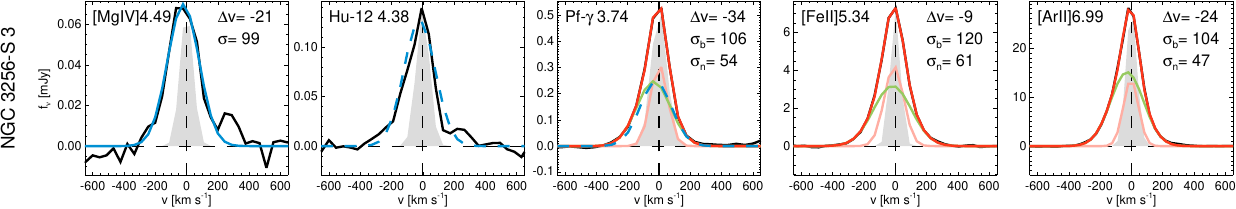}
\includegraphics[width=0.97\textwidth]{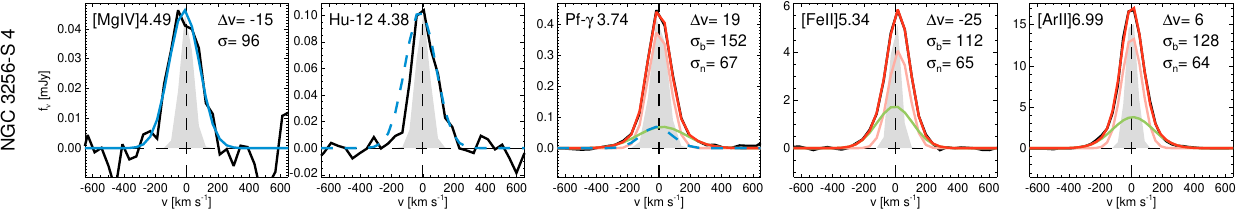}
\includegraphics[width=0.97\textwidth]{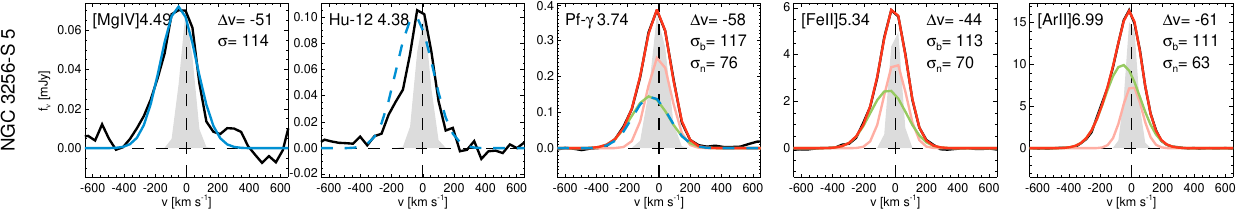}
\includegraphics[width=0.97\textwidth]{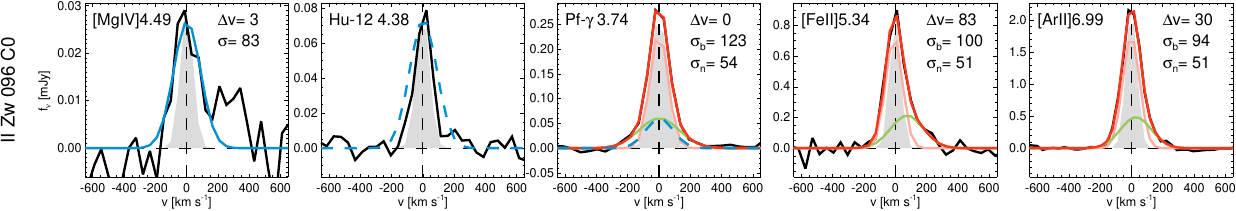}
\includegraphics[width=0.97\textwidth]{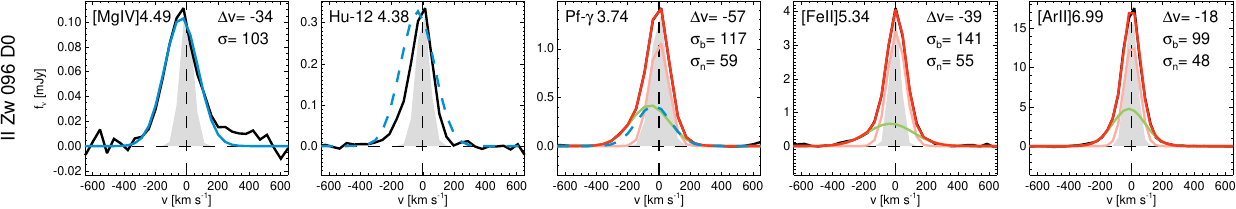}
\includegraphics[width=0.97\textwidth]{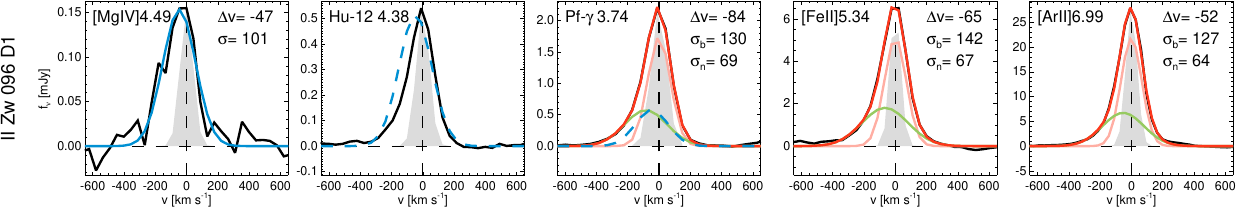}
\caption{Continued.}
\end{figure}

\begin{figure}[h]
\addtocounter{figure}{-1}
\centering
\includegraphics[width=0.97\textwidth]{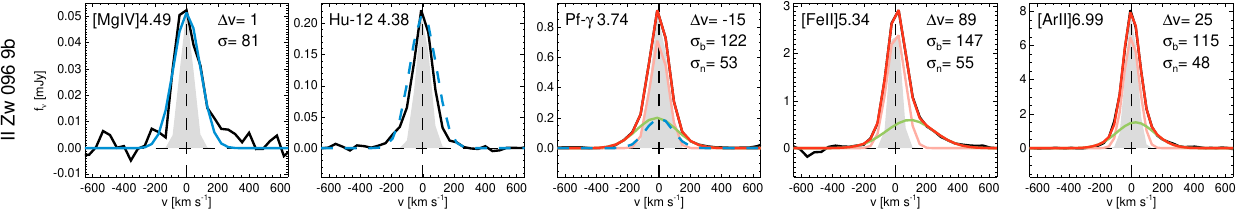}
\includegraphics[width=0.97\textwidth]{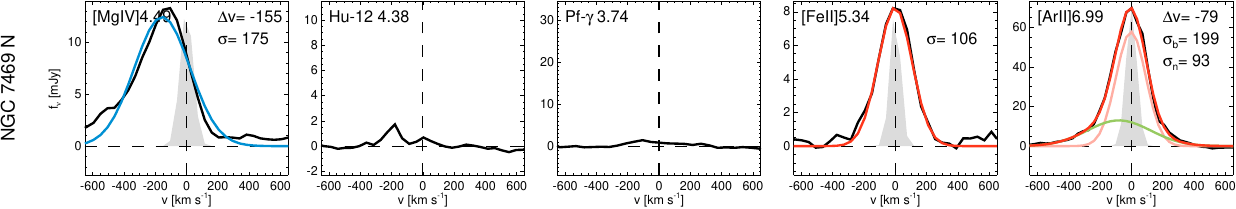}
\includegraphics[width=0.97\textwidth]{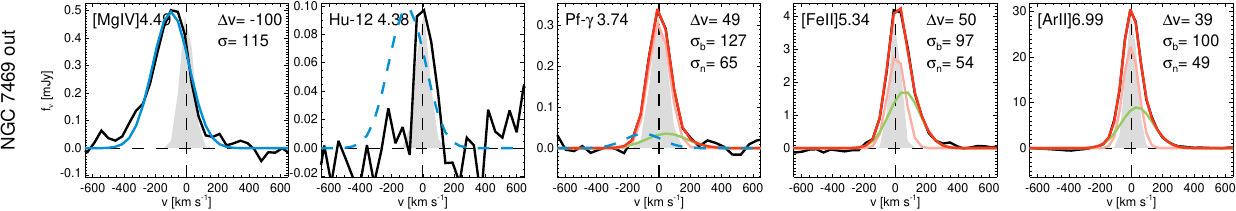}
\includegraphics[width=0.97\textwidth]{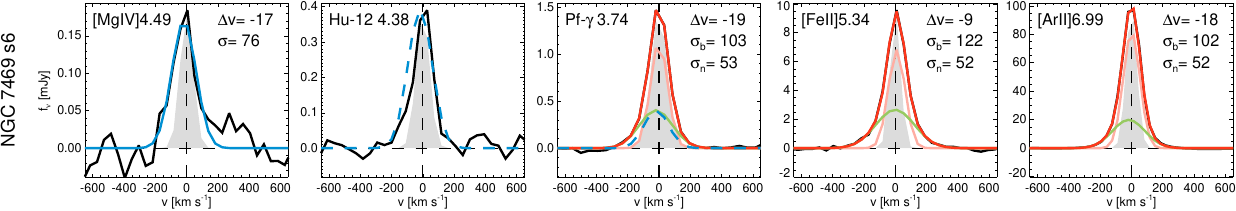}
\includegraphics[width=0.97\textwidth]{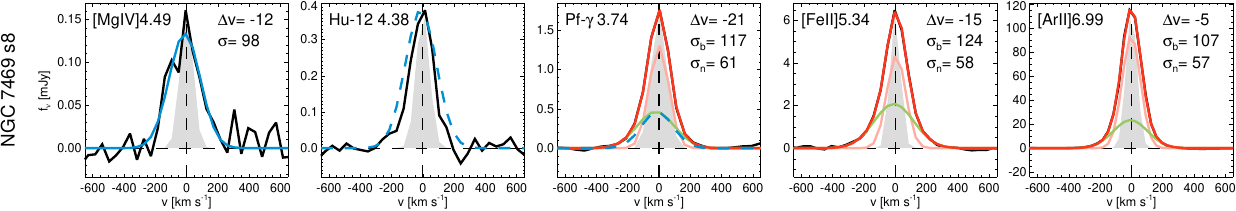}
\includegraphics[width=0.97\textwidth]{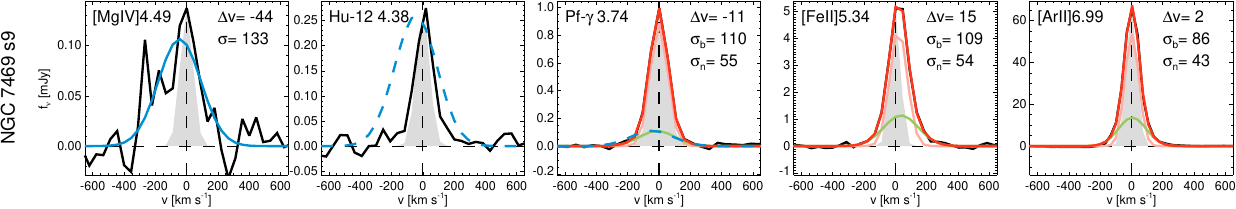}
\caption{Continued.}
\end{figure}

\clearpage
\twocolumn

\section{Relation between the luminosity and kinematics of \Mgiv\ and H$_2$ S(8), \Feii, and \Arii}

\begin{figure}
\centering
\includegraphics[width=0.4\textwidth]{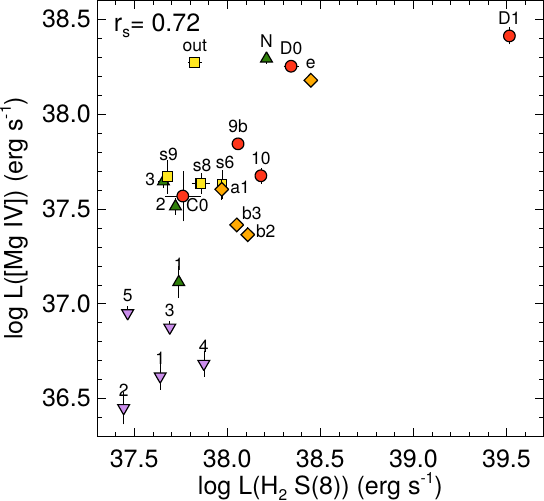}\\[0.3cm]
\includegraphics[width=0.4\textwidth]{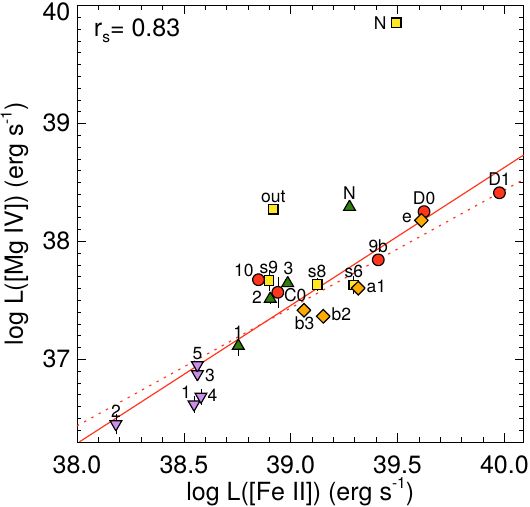}
\caption{\Mgiv\ vs. H$_2$ S(8) {(top) and vs. \Feii\ (bottom)} luminosities. The symbols are as in Fig.~\ref{fig_flux_relation_hu12}.
{The solid red line is the best fit ($\log y = -8.20 + 1.17 \log x$) and the dashed red line the best linear fit ($\log y = (-1.56 \pm 0.25) + \log x$).}
\label{fig_flux_relation_h2}}
\end{figure}

\begin{figure}
\centering
\includegraphics[width=0.4\textwidth]{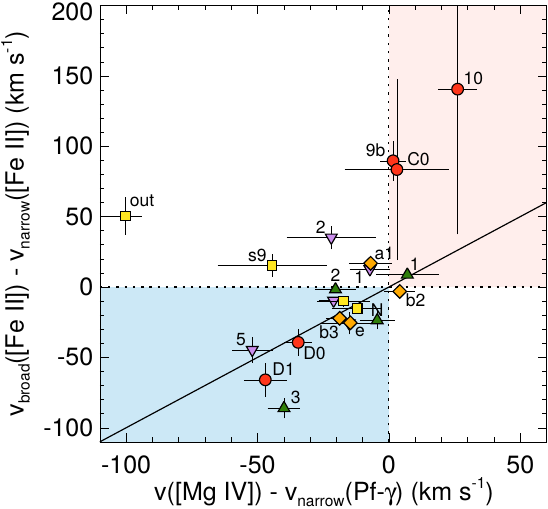}\\[0.3cm]
\includegraphics[width=0.4\textwidth]{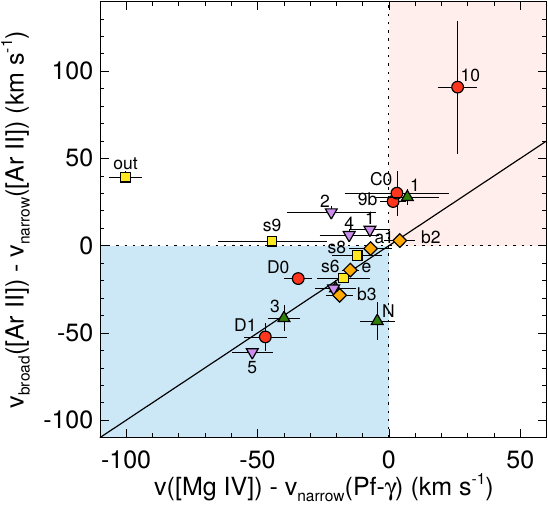}
\caption{Same as the bottom panel of Fig.~\ref{fig_kin_relations} but for \Feii\ (top) and \Arii\ (bottom). \label{fig_kin_relations_fear}}
\end{figure}

Fig.~\ref{fig_flux_relation_h2} shows the relation between the \Mgiv\ and {the H$_2$ S(8) and \Feii\  5.34\micron} luminosities. {We scaled the MRS spectra to match the NIRSpec continuum at $\sim$5\micron. The scaling factors are between 0.7 and 1.3. We did not correct for the differential extinction affecting these lines.}
Fig.~\ref{fig_kin_relations_fear} presents the comparison between the \Mgiv\ line kinematics observed with JWST\slash NIRSpec and the longer wavelength lines (\Feii\ and \Arii) observed with JWST\slash MRS.

\section{Star-formation, AGN, and shock models}\label{apx:models}

\begin{figure}
\centering
\includegraphics[width=0.4\textwidth]{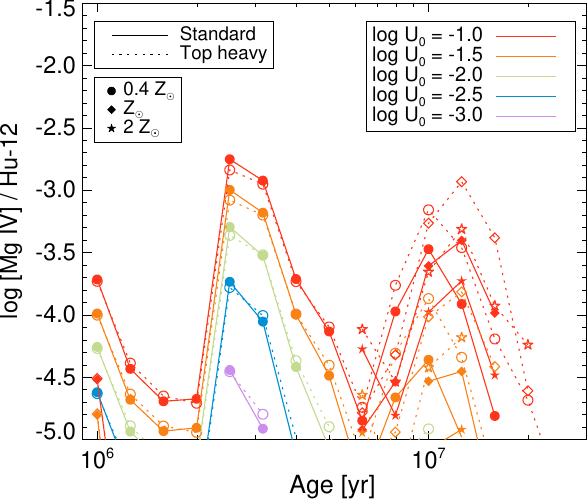}
\caption{\Mgiv\slash Hu-12 ratio for an instantaneous burst of star-formation as function of the stellar age.
The circles, diamonds, and stars mark the ratios for 0.4$Z_\odot$, $Z_\odot$, and 2$Z_\odot$. The filled (empty) symbols connected by a solid (dashed) line corresponds to a standard (top heavy) IMF. The grid with $n_{\rm H}=10^2$\,cm$^{-3}$ is plotted. The color of the lines indicates the ionization parameter, $U_0$.
\label{fig_sf_model}}
\end{figure}

\begin{figure*}[t]
\centering
\includegraphics[width=0.75\textwidth]{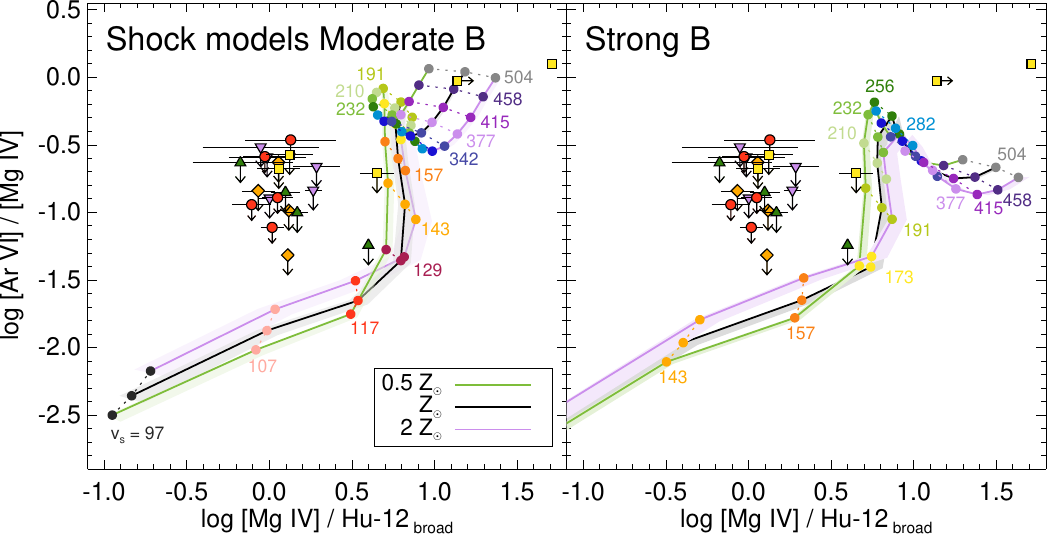}
\caption{Same as the right panel of Fig.~\ref{fig_model_agn_shock} but for the moderate (left) and strong (right) magnetic field cases.
\label{fig_shock_mag}}
\end{figure*}

In this section, we present a more detailed description of the grids of models used in Sect.~\ref{s:models}. For the star-formation and AGN photoionization models, we used \textsc{Cloudy} version 23.01 \citep{Chatzikos2023} and mostly followed \citet{Pereira2017FIR} to create the grids. The main differences are the updated SEDs used for the incident radiation field (see below). For the shock models, we used the \textsc{MAPPINGS V} code version 5.2.0 following \citet{Sutherland2017}.

\subsection{Star-formation models}\label{apx:mod_sf}

We used a constant density slab model. The indicent radiation field is obtained from the BPASS library (version 2.2; \citealt{Eldridge2017}). This library includes the effect of stripped-envelope stars in binary systems, which increase the hardness of the spectrum.
We assumed an instantaneous burst of star-formation with ages between 10$^6$ to 10$^8$ with 0.1\,dex steps. We considered three stellar metallicities (0.4\,\Zsun, \Zsun, and 2\,\Zsun), and two initial mass functions (IMFs), both with an upper stellar mass limit of 300\,\Msun: a standard \citet{Kroupa2001} IMF and a top-heavy IMF with a power-law exponent of $-2$ between 0.5 and 300\,\Msun.
The initial gas phase abundance was matched to the stellar metallicity and the metals depleted using the ``ISM\_CloudyClassic.dpl'' table. The gas-to-dust mass ratio were adjusted similar to \citet{Pereira2017FIR}.
We varied the gas volume density, $n_{\rm H}$, between 10$^2$ and 10$^4$\,cm$^{-3}$ in 1\,dex steps and the ionization parameters between $\log U$=$-6.75$ and $-1.0$ in 0.25\,dex steps. The models are stopped when the temperature drops below 1000\,K or the
H$^+$ abundance is below 0.1\%.

In order to trace the temporal evolution of the line ratios, we need to define how $\log U$ varies with the age of the stellar population. $\log U$ is proportional to the rate of ionizing photons, $Q({\rm H})$. Thus, 
{similar to \citet{Rigby2004}}, we defined, for every combination of IMF and stellar metallicity, $\log U_0$ as the ionization parameter when $Q({\rm H})$ is maximum ($Q({\rm H})_{\rm 0}$; typically at ages of $\sim$1--2\,Myr), and scaled $\log U_{\rm age}$ relative to $\log U_0$ using the ratio $Q({\rm H})$(age)/$Q({\rm H})_{\rm 0}$.
Then, the line emission at each stellar age is interpolated from the grid using $\log U_{\rm age}$.

Figure~\ref{fig_sf_model} shows the \Mgiv\slash Hu-12 ratio as function of the stellar age for $\log U_0$ between $-1.0$ and $-3.0$. Only the $n_{\rm H}=10^2$\,cm$^{-3}$ ratios are presented since the variation of this ratio in the explored $n_{\rm H}$ range (10$^2$--10$^{4}$\,cm$^{-3}$) is small, $<$30\,\%.

\subsection{AGN photoionization models}\label{apx:mod_agn}

Similar to the star-formation models (Sect.~\ref{apx:mod_sf}), we assumed a constant density slab. For the incident radiation, we used the three SEDs derived by \citet{Jin2012} for low, medium, and high Eddington ratios. These SEDs represent a more realistic characterization of the SED of AGN than a power-law (e.g., \citealt{Ferland2020}).
We considered three gas phase abundances (0.4\,\Zsun, \Zsun, and 2\,\Zsun) which were depleted using the ``ISM\_CloudyClassic.dpl'' table and their gas-to-dust mass ratio were adjusted (see Sect.~\ref{apx:mod_sf}).
We varied $n_{\rm H}$ between 10$^2$ and 10$^5$\,cm$^{-3}$ in 0.5\,dex steps and $\log U$ between $-4.0$ and $-1.75$ in 0.25\,dex steps.
Line ratios derived from these grids are shown in the left panel of Fig.~\ref{fig_model_agn_shock}.

\subsection{Shock models}\label{apx:mod_shock}

We used \textsc{MAPPINGS V} code version 5.2.0 to create a grid of shock models \citet{Sutherland2017}.
This grid includes several improvements compared to the models presented by \citet{Allen2008}.
For instance, the fully time-dependent solution for the photoionized precursor is computed for the first time (see \citealt{Sutherland2017} for details).

The effect of dust is not taken into account by the \textsc{MAPPINGS V} shock code. This is justified for the high-temperature post-shock gas where dust grains can be effectively destroyed (see \citealt{Allen2008}). However, dust will be present in the photoionized precursor. This is particularly important since Mg is known to condense on silicate grains (e.g., \citealt{Rogantini2020}). As a first order correction, we reduced the gas phase abundances of metals in the precursor using the default \textsc{cloudy} depletion factors (0.2 for Mg).
We also updated the atomic parameters relevant for the emission lines of interest (e.g., \Mgiv\ and \Arvi) to match those used in \textsc{Cloudy} version 23.01. We used atomic data from the CHIANTI database version 10 \citep{DelZanna2021} and the Stout database from \textsc{Cloudy} \citep{Lykins2015}.

We followed \citet{Sutherland2017} to create the grids varying four input parameters: the gas metallicity, $Z_{\rm gas}$; the shock velocity, $v_{\rm s}$; the ram pressure parameter, $R=$\hbox{$(n_{\rm H}\slash 1\,{\rm cm^{-3}})$} $\times $\hbox{$(v_{\rm s}\slash {\rm km\,s^{-1}})^2$}; and the magnetic to ram pressure ratio, $\eta_{\rm M}=B^2\slash (4\pi\rho v_{\rm s}^2)$, where $B$ is the magnetic field and $\rho$ the density.
For $Z_{\rm gas}$, we used 0.5$Z_\odot$, $Z_\odot$, and 2$Z_\odot$ and for $v_{\rm s}$ we used a logarithmic grid between 80 and 500\,km\,s$^{-1}$ with $\sim$0.04\,dex steps. We used three pressures, $R=10^4$, 10$^6$, and 10$^8$, which are equivalent to densities of $n_{\rm H}=1, 100$, and 10$^4$\,cm$^{-3}$ at 100\,km\,s$^{-1}$. For the magnetic field, we used $\eta_{\rm M}=10^{-4}, 10^{-2}$, and 10$^{-1}$, which correspond to the ``standard'', ``moderate'', and ``strong'' magnetic cases identified in \citet{Sutherland2017} and are equivalent to $B=$5.4, 54, and 170\,$\mu$G for $R=10^6$. The standard magnetic case is presented in the right panel of Fig.~\ref{fig_model_agn_shock} and the moderate and strong in Fig.~\ref{fig_shock_mag}. The main effect of the magnetic field on the \Mgiv\slash Hu-12 ratios is that
that a higher $v_{\rm s}$ is needed to obtain the same ratio because more energy is needed to compress the gas when the magnetic field is stronger.

\subsection{\Mgiv\ luminosity from SNe}\label{apx:mod_shocks_energy}

We estimated the expected \Mgiv\ luminosity produced by SNe shocks for a constant star-formation rate, traced by H recombination lines, by combining the shock models and the stellar population evolution models from the BPASS library.

Assuming Case B conditions, the H recombination lines can be used to measure $Q({\rm H})$. For temperatures between 5\,000 and 10\,000\,K, the H$\alpha$\slash Hu-12 ratio is 1250--1820 \citep{Storey1995} and $Q({\rm H})$(s$^{-1}$)$ = (0.85-1.41)\times10^{15} L$(Hu-12)(erg\,s$^{-1}$).

According to the BPASS stellar population models, for metallicities between 0.4\,\Zsun\ and 2\Zsun, and assuming a constant star-formation rate, $\log (Q$(H)(s$^{-1}$)\slash $\dot{E}_{\rm SN}$(erg\,s$^{-1}$))=11.72--12.10, where $\dot{E}_{\rm SN}$ is the power released by SNe.
This ratio is sensitive to the assumed energy released per SN event (10$^{51}$\,erg), and also to the IMF (e.g., \citealt{Zapartas2017}). For this calculation, we used the \citet{Kroupa2001} IMF and a top-heavy IMF (see Sect.~\ref{apx:mod_sf}).

\begin{figure}[h]
\centering
\includegraphics[width=0.4\textwidth]{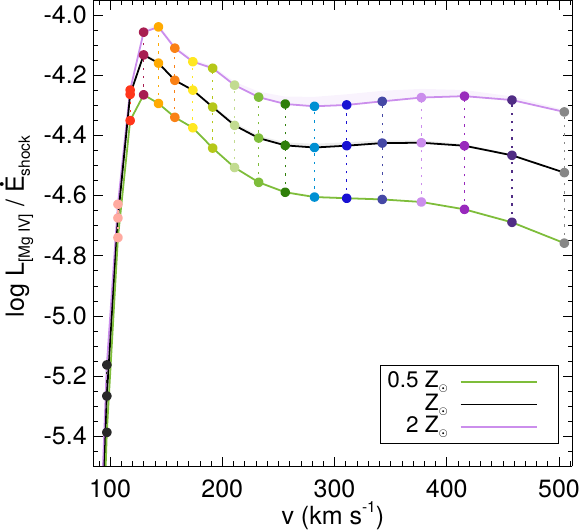}
\caption{Ratio between the \Mgiv\ luminosity and the rate of mechanical energy flux across the shock, $\dot{E}_{\rm shock}$. The symbols are as in the right panel of Fig.~\ref{fig_model_agn_shock}.
\label{fig_shock_energy}}
\end{figure}

For these metallicities, shock models (Sect.~\ref{apx:mod_shock}) predict that \Mgiv\ is emitted more efficiently at $v_{\rm s}$$\sim$110--160\,km\,s$^{-1}$. These $v_{\rm s}$ are also comparable to the observed dispersion of the \Mgiv\ line profiles. Therefore, in this $v_{\rm s}$ range, $\log L$(\Mgiv)\slash $\dot{E}_{\rm shock}$, where $\dot{E}_{\rm shock}=0.5\rho v_{\rm s}^3$ is the rate of mechanical energy flux across the shock, is between --4.25 and --4.05 (Fig.~\ref{fig_shock_energy}). {We note that this fraction is very small compared to the $\sim$0.2\%\ shock energy that is released by some \Feii\ transitions (e.g., \citealt{Mouri2000}).} 

Combining these three relations, matching the abundances of the stellar population and the shock models, and taking into account the Hu-12 emission from the shock (right panel of Fig.\ref{fig_model_agn_shock}), the predicted $\log L$(\Mgiv)\slash $L$(Hu-12) ratio in a region with constant star-formation rate and shocks produced by SNe would be --1.4 and --0.5.

\end{document}